\documentclass[preprint,authoryear,12pt]{elsarticle}

\usepackage{epsfig}

\usepackage{amssymb}

\journal{Advances in Space Research}

\begin{document}

\begin{frontmatter}



\title{RELATIVISTIC POSITIONING AND SAGNAC-LIKE MEASUREMENTS FOR
FUNDAMENTAL PHYSICS IN SPACE}


\author{Angelo Tartaglia }
\address{INAF-OATo and Politecnico di Torino, Corso Duca degli Abruzzi 24, 10129 Torino, Italy}
\ead{angelo.tartaglia@polito.it}




\begin{abstract}

The paper concerns the use of satellites of the Galileo constellation for relativistic positioning and for
measurements of the gravito-magnetic effects induced by the angular momentum both of the Earth and of the
dark halo of the Milky Way. The experimental approach is based on the generalized Sagnac effect, induced both
by the rotation of the device and the fact that the observer is located within the gravitational field of a
spinning mass. Among the possible sources there is also the angular momentum of the dark halo of the Milky
Way. Time modulation of the expected signal would facilitate its disentanglement from the other
contributions. The modulation could be obtained using satellites located on different orbital planes.

\end{abstract}


\end{frontmatter}

\parindent=0.5 cm

\section{Introduction}

The opportunity to use the Galileos for positioning purposes with a method differing from the typical GPS
one and incorporating relativity rather than treating its effects as corrections to the data has been described in
various papers, see for example \citep{TA10}\citep{TRC11}. The use of that method for accurately determining the orbits of the Galileo GSAT-5 and 6 satellites, accidentally ended up in eccentric orbits, was proposed in the previous edition of the present conference \citep{TA17}.

In general, the need for an accurate fully relativistic positioning is compelling when dealing with space missions beyond the terrestrial environment, as it is the case, for instance, of the ASTROD projects where an accurate use of laser ranging, together with orbit determination based on post Newtonian development of General Relativity (GR) is envisaged. See, for instance Ref.s \citep{app09,brax12}.

\begin{figure}[h]
\label{figure1}
\begin{center}
\includegraphics*[width=12cm,angle=0]{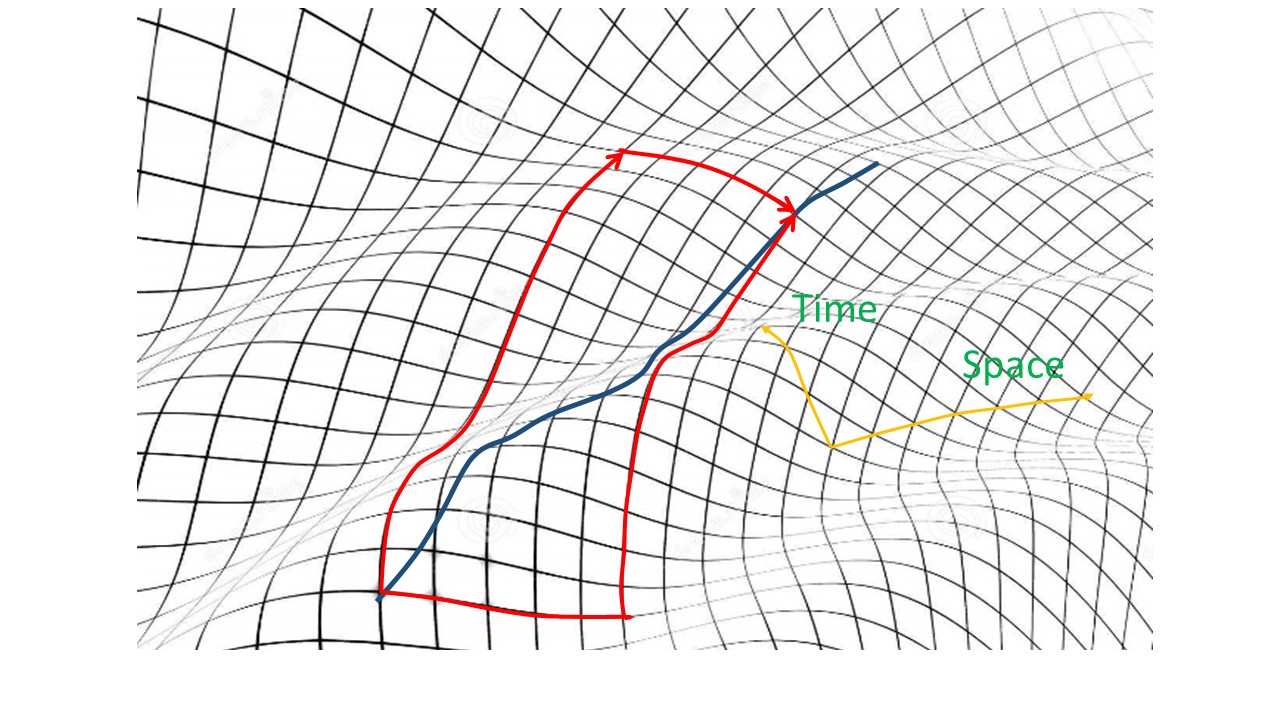}
\end{center}
\caption{1+1 dimensional representation of the spacetime
continuum. The lattice is made of light rays.}
\end{figure}

For us here, the starting remark is that space-time is a geometric four-dimensional Riemannian manifold. The problem of
positioning in space and time in such manifold can be approached in a way similar to topography in ordinary
two or three dimensions. One starts from a set of four (as the dimensions of the manifold) stations, whose
positions in space-time form a reference basis. If we think of four satellites orbiting the Earth, we have a
configuration evolving in space, according to a known pattern; equipping each station with an accurate clock,
we are able to "draw" the world-lines of our beacons. If each satellite emits regular pulses, a traveller receiving
them, just recording the arrival time sequences, is enabled, by the use of a simple algorithm, to reconstruct
its coordinates with respect to the basis \citep{TRC11}. The "drawing" of the world-line of the receiver incorporates the curvature of
the manifold, i.e. it accounts for the effect of gravity. The idea is sketched in Fig. \ref{figure1}, where the blue line is the
world-line of an observer; red lines together form a closed loop (as viewed by the observer) travelled by
light in opposite directions. Yellow arrows represent time (left) and space (right) directions at a given position
of the lattice.

In fact, the simple image presented in Fig. \ref{figure1} contains also the essence of the generalized Sagnac method. A
real configuration would not be so simple, requiring not less than three (in the figure they are only two) mirrors
and not less than two space dimensions (in the figure there is just one) in order to form a closed polygon in the
reference frame moving with the observer. The relevant feature is that travelling along the same space path in
opposite directions requires different times of flight as measured by the observer. This asymmetry is precisely
the experimental target of the proposals we are describing.

\section{GRAVITO-MAGNETISM AND SAGNAC}
\label{Section 2}

The starting point of our reasoning can be the general
form of the line element in a given space-time:

  \begin{equation}
    \label{eq:1}
 ds^2=g_{00}c^2dt^2+g_{ij}dx^i dx^j +2g_{0i}cdtdx^i
  \end{equation}

Notations and conventions are standard in GR. The coordinates are arbitrary.

If we wish to describe an electromagnetic (EM) signal, its line element is null, i.e. $ds = 0$. We may then use Eq. (\ref{eq:1})
 to work out the coordinated time of flight (tof) along a length $dl$ of a given space path:

 \begin{equation}
 \label{eq:2}
 dt=-\frac{g_{0i}}{cg_{00}}\frac{dx^i}{dl}dl+\frac{\sqrt{(g_{0i}dx^i/dl)^2-g_{00}g_{ij}(dx^i/dl)(dx^j/dl)}}{cg_{00}}dl
 \end{equation}

 The choice of the $+$ sign in front of the second term insures that the propagation is always toward the future.
We may now integrate (\ref{eq:2}) along a closed path (of course we need some device - mirrors or waveguide - in order
to oblige light to move that way) in order to find the round trip tof. If we do so in one sense, then repeat the
operation in the opposite sense, all $dx^i$'s change their sign. The final result is that the first term in (\ref{eq:2}) reverses
its sign, but the second term remains as it is.

Let us subtract the two integrals from one another and
we end up with the tof asymmetry:

\begin{equation}
\label{eq:3}
\delta\tau=-\frac{2}{c}\sqrt{g_{00}}\oint\frac{g_{0i}}{g_{00}}\frac{dx^i}{dl}dl
\end{equation}

The symbol $\tau$ is used to denote the proper time of the observer, i.e. the time measured by the observer in its
rest-frame. This is the reason why on the right, in front of the integral, a $\sqrt{g_{00}}$ is found: it accounts for the
gravitational potential at the position of the observer.

\subsection{Coordinate and physical effects}
\label{Subsection 2.1}

As we see, looking at Eq. (\ref{eq:3}), the tof asymmetry may be nonzero only if $g_{0i}\neq 0$. The mixed time-space terms
of the metric tensor express, in general, a state of motion of the observer with respect to the chosen reference
frame. If the motion is purely inertial a simple coordinate transformation can bring all $g_{0i}$'s to zero
everywhere: we are facing a simple coordinate effect.

More interesting is the case when, even in a flat spacetime (no gravity), the observer (because of any
mechanical constraint) is \textit{rotating} with respect to the inertial axes of the frame: this is the classical Sagnac
effect (even though in relativistic notation) which is exploited on airplanes or other vehicles for measuring
the angular velocity. We may call it a \textit{kinematical} effect.

Finally, when a gravitational field is present and the source is spinning, it is not possible, by a change of
reference, to eliminate \textit{globally} the mixed time-space elements of the metric tensor (it is always possible to do
so \textit{locally}). This is the situation when, in weak field condition, people speak of gravito-magnetic (GM) field
and the $g_{0i}$'s may be interpreted as the components of a potential three-vector of that field \citep{RT2}.

\section{WHAT CAN BE MEASURED}
\label{Section 3}

According to the effects mentioned in Sect. \ref{Section 2}, in a general experiment we may expect to find a kinematical
Sagnac term leading to a measured asymmetry given by the formula \citep{RT2}

\begin{equation}
\label{eq:4}
\delta\tau_S\simeq 4\frac{S}{c^2}\Omega
\end{equation}

$S$ is the area contoured by the path of light; $\Omega$ is the angular velocity of the apparatus with respect to the
"fixed stars".

$\delta\tau_S$ is usually dominant, but if the observer is in the space-time surrounding a spinning mass whose angular
momentum be $\bar{J}$ we may use the approximated line element (weak field approximation and polar space coordinates):

\begin{equation}
\label{eq:5}
ds^2\simeq \left(1-2\frac{m}{r}\right)c^2dt^2-\left(1+2\frac{m}{r}\right)dr^2-r^2d\theta^2-r^2\sin^2\theta d\phi^2+4\frac{j}{r^2}\sin^2\theta rd\phi cdt
\end{equation}

where, being $M$ the mass of the source (assumed to be spherical), it is

\begin{equation}
\label{eq:6}
m=G\frac{M}{c^2}\qquad \textrm{and} \qquad j=G\frac{J}{c^3}
\end{equation}

The smallest term in (\ref{eq:5}) is the one containing $j$ and that will be the border for any further approximated
calculation.

In the case of the Earth it is
\begin{equation}
\label{eq.7}
m=4.43\times10^{-3}\, \textrm{m}\qquad \textrm{and} \qquad j=1.75\times10^{-2}\, \textrm{m}^2
\end{equation}

As I have already written, the time-space term of the metric may here be interpreted as the only non-zero
component of a potential three-vector $\bar{h}$:

\begin{equation}
\label{eq.8}
h_{\phi}=2\frac{j}{r^2}\sin{\theta}
\end{equation}

which leads to the GM field

\begin{equation}
\label{eq.9}
\bar{B}_g=\bar{\nabla}\wedge\bar{h}=\frac{2}{r^3}\left[\bar{j}-3(\bar{j}\cdot\hat{u}_r)\hat{u}_r\right]
\end{equation}

\section{USING GALILEO}
\label{Section 4}

Let us consider three appropriately chosen satellites on the same orbital plane: they form a "rigid" equilateral
triangle having the Earth at the center, like in Fig. 2.

\begin{figure}[h]
\label{fig:2}
\begin{center}
\includegraphics*[width=10cm,angle=0]{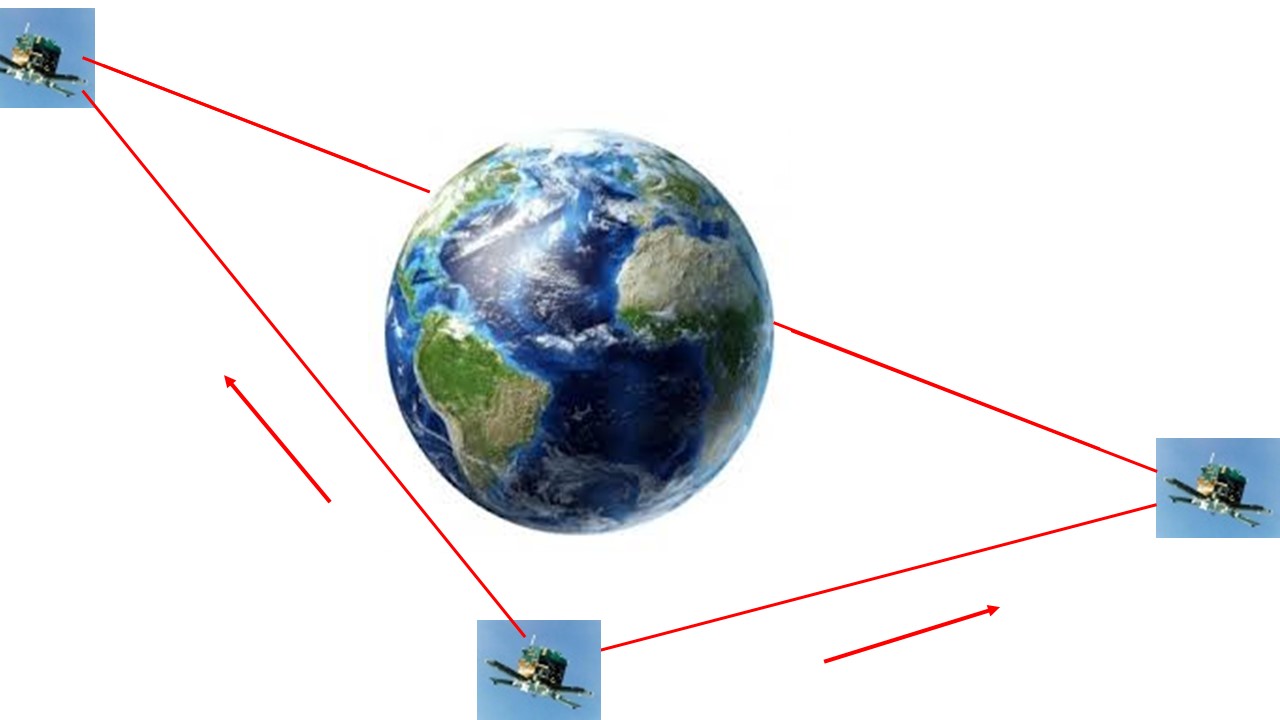}
\end{center}
\caption{Three Galileo satellites exchanging
electromagnetic signals along an equilateral triangle.}
\end{figure}

If one of the satellites acts as the main station, it can send EM signals both ways toward the others, which in turn
resend them along the sides of the triangle. The main station, as orbiting observer, measures the tof difference.
The appropriate reference frame is now co-rotating with the satellites so the relevant line element must be
deduced from Eq. (\ref{eq:5}), applying a rotation and a boost along the tangent to the orbit; the result is

\begin{eqnarray}
\label{eq.10}
ds^2\simeq\left(1-2\frac{m}{r}-\frac{\Omega^2r^2}{c^2}\sin^2{\theta}\right)c^2dt^2 \nonumber \\
-\left(1+2\frac{m}{r}\right)dr^2-r^2d\theta^2-r^2\sin^2{\theta}d\phi^2 \\
+2\left(2\frac{j}{r^2}+\frac{\Omega r}{c}-2\frac{\Omega m}{c}\right)r\sin^2{\theta}d\phi c dt \nonumber
\end{eqnarray}

In the brackets of the last line we recognize, in the order, the GM term produced by the angular momentum of the
Earth, the kinematical Sagnac term and the de Sitter (or geodetic) term, coupling the orbital rotation with the
gravitational potential of the planet.

Being the satellites in free fall, the angular velocity, at our approximation level, is Keplerian, i.e.

\begin{equation}
\label{eq.11}
\Omega=\frac{c}{r}\sqrt{\frac{m}{r}}
\end{equation}

For simplicity let us imagine that our triangle (unlike the real Galileo satellites) is in the equatorial plane.
Using (\ref{eq.11}) the line element then becomes:

\begin{eqnarray}
\label{eq.12}
ds^2\simeq \left(1-3\frac{m}{r}\right)c^2dt^2-\left(1+2\frac{m}{r}\right)dr^2-r^2d\phi^2 \nonumber \\
+2\left(2\frac{j}{r^2}+\sqrt{\frac{m}{r}}-2\left(\frac{m}{r}\right)^{3/2}\right)r d\phi cdt
\end{eqnarray}

From (\ref{eq.12}) and (\ref{eq:3}) we get:

\begin{equation}
\label{eq.13}
\delta \tau \simeq -\frac{2}{c}\oint\left(2\frac{j}{r}+\sqrt{mr}\right)d\phi
\end{equation}

\subsection{Expected values}
\label{Subsection 4.1}

We may now apply (\ref{eq.13}) to our triangle. Still thinking to be in the equatorial plane (the final result will give us an
idea of the expected signal modulo a factor proportional to the cosine of the inclination of the orbital plane), the
equation of the sides is:

\begin{equation}
\label{eq.14}
r(\phi)=\frac{b}{\cos{(\phi-\phi_H)}}
\end{equation}

where $b$ is the closest approach distance from the Earth and $\phi_H$ is the azimuth in the middle of the side. The geometry is sketched in Fig. 3:
\begin{figure}[h]
\label{fig:3}
\begin{center}
\includegraphics*[width=12cm,angle=0]{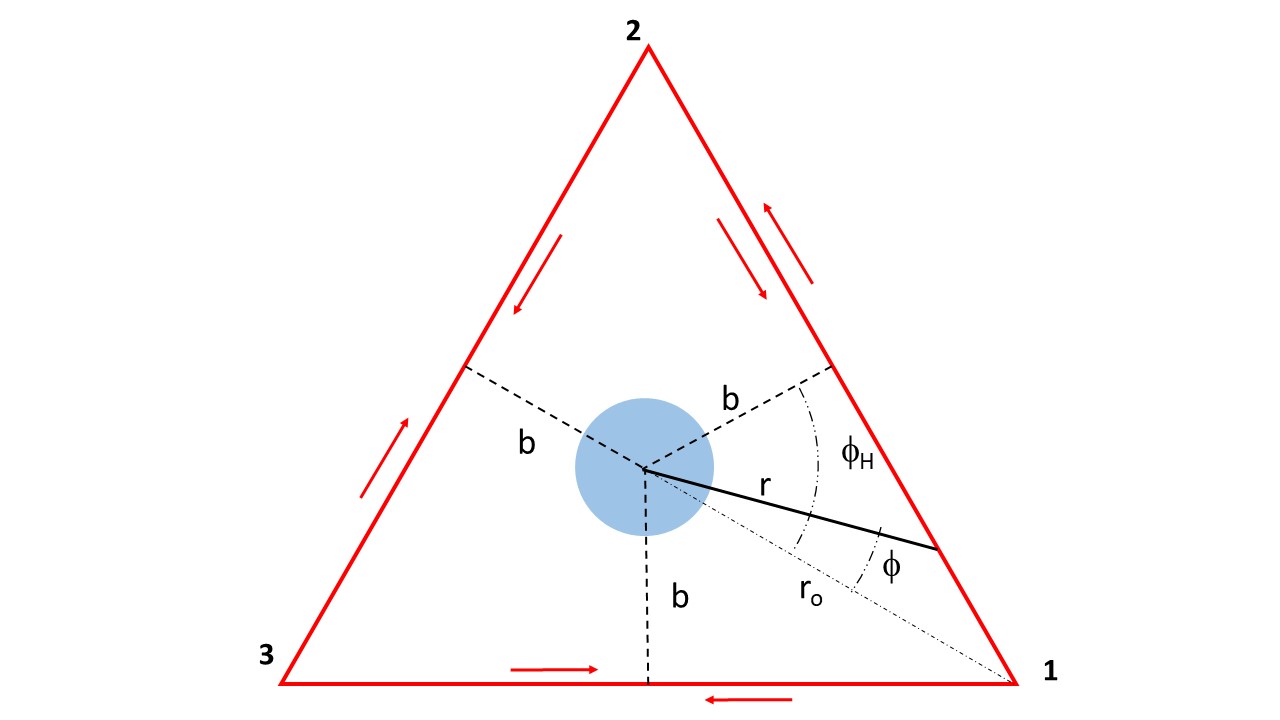}
\end{center}
\caption{Sketchy configuration of three satellites in the equatorial plane at the vertices of an equilateral triangle. The earth is the blue disk in the center. Red lines represent EM rays; the parameters are described in the text.}
\end{figure}

We are assuming that the EM trajectories are indeed straight, which is not literally true. The Earth's gravity produces
a lensing, but: a) the phenomenon is absolutely negligible at our scale; b) in any case the effect of the
curvature is symmetric, then not affecting the tof asymmetry we are looking for.

Under these conditions the expected effect due to the angular momentum of the Earth is \citep{RT19}:

\begin{equation}
\label{eq.15}
|\delta \tau_j|\simeq 24\sqrt{3}\frac{j}{cr_o}
\end{equation}

Here $r_o$ is the radius of the (circular) orbit of the satellites.

If we apply (\ref{eq.15}) to geostationary orbits ($r_o = 42.164 \times 10^6\, \textrm{m}$) the numerical value is:

\begin{equation}
\label{eq.16}
|\delta\tau_j|_{gs}\simeq 5.75 \times 10^{-17}\,\textrm{s}
\end{equation}

If instead we use the radius of the orbits of the Galileos ($r_o = 29.600 \times 10^6 \,\textrm{m}$) it is:

\begin{equation}
\label{eq:17}
|\delta \tau_j|_G \simeq 8.19\times10^{-17}\,\textrm{s}
\end{equation}

For true Galileo orbits we must consider that the orbital plane is at an angle $\alpha = 56^o $ with respect to the equator,
so that in the evaluation of the integral along the orbit
we must keep a factor
\begin{equation}
\label{eq:18}
\sin^2{\theta}=\frac{1}{1+\tan^2{\alpha}\cos^2{\phi}}
\end{equation}

The final result for the GM contribution is
\begin{center}
$|\delta \tau_j|_G \simeq 7.55\times 10^{-17}$\,s
\end{center}

These numbers compete with the other effect in Eq. (\ref{eq.13}) which is the much bigger Sagnac effect. For the radius
of geostationary orbits, it is (numerical evaluation)
\begin{center}
$|\delta\tau_j|_{gs}\simeq 9.91\times 10^{-6}$\,s
\end{center}

For Galileo orbits it is:
\begin{center}
$|\delta \tau_j|_G \simeq 8.30\times 10^{-6}$\,s
\end{center}
A necessary condition for spotting the effect (\ref{eq.15}) is of course to be able to remove from the signal the
absolutely dominant kinematical Sagnac.

\section{GALACTIC DARK HALO}
\label{Section 5}

Current scientific wisdom, in order to justify the observed peripheral velocities of stars, assumes that
most spiral galaxies (including ours) are immersed in a huge dark matter halo, bigger than the galaxy \citep{WT18}. Now, if that halo actually exists, in a more or less stationary situation it is reasonable to expect that it also rotates together with the visible
galaxy. If so, and considering that the estimated mass of the halo is many times larger than that of the visible
galaxy \citep{KSLB14}, it should also possess an important angular momentum. Final induction: according to GR the
(rotating) dark halo produces a GM field and it would be worth looking around for its effects.

The axial symmetry of the Milky Way tells us that the galactic GM field, $\bar{B}_{MW}$, should have a dipolar
configuration. In the galactic plane (where the solar system is approximately located) $\bar{B}_{MW}$ is then
perpendicular to that plane.

Of course the intensity of the field will depend on the distance from the axis of the Milky Way, which is
located at $R_G \simeq 2.3\times10^{20}$\,m away from us \citep{FA14}. The distance periodically changes
while the Earth rotates around the Sun over a range given by the diameter of the Earth's orbit, $D_E \sim 3\times10^{11}$\,
m; the orbital plane is also at an angle with respect to the galactic plane, but this is not important at the moment.
Reasonably we are allowed to say that the relative change of the galactic GM field in the inner solar system
is
\begin{equation}
\label{eq:17}
\frac{\delta B_{MW}}{B_{MW}}\propto \frac{D_E}{R_G}\sim 10^{-9}
\end{equation}

In practice we are allowed to treat $\bar{B}_{MW}$ as constant throughout the terrestrial orbit.

\subsection{Estimating the measurable effect}
\label{Subsection 5.1}

In the case of the effect of the dark halo of the Milky Way in correspondence of the solar system we cannot start from the line element (\ref{eq:5}), because now the observer is inside the mass distribution of the source and not
outside. Our basic formula will then be Eq. (\ref{eq:3}), however exploiting the axial symmetry of the distribution. Using
Stokes' theorem and recalling the practical uniformity of the galactic GM field perceived by a terrestrial
observer, the formula for the tof asymmetry induced by the rotating halo is reduced to the calculation of the flux
of a constant vector. The result is very simple:

\begin{equation}
\label{eq:19}
\delta \tau_{MW}\simeq \frac{2}{c}\oint \bar{h}\cdot d\bar{l}=\frac{2}{c}\int \bar{B}_{MW}\cdot \hat{u}_n dS\simeq \frac{2}{c}B_{MW}S\cos\beta
\end{equation}

The flux is calculated across the area $S$, contoured by the loop along which the EM signals travel; $\hat{u}_n$ is the
unit vector perpendicular to that area, assumed to be in a plane; $\beta$ is the angle between the axis of the Milky Way
and the direction indicated by $\hat{u}_n$.

From (\ref{eq:19}) we may express the detectable galactic GM field as a function of the achievable sensitivity, $\delta \tau_{min}$, of
the measurement:

\begin{equation}
\label{eq:20}
B_{MW}\geq \frac{c\delta\tau_{min}}{2S\cos\beta}
\end{equation}

In the case of the triangle made of Galileo satellites it would be $S \simeq 1.14\times 10^{15}$\,m$^2$ and $\beta \simeq 62.6^o$. As for the
inclination, the number accounts for the orientation of the ecliptic plane with respect to the galactic plane; it
should then be combined with the inclination of the axis of the Earth with respect to the celestial North ($23.45^o$)
and that of the orbital plane of the Galileos with respect to the equator of our planet ($56^o$): these details are
however not important for the present purpose.

Just to have an idea about numbers, let us consider a GM field intensity as high as the one of the Earth measured on the surface of the planet: $B_{E}\sim 10^{-23}\,\textrm{m}^{-1}$.\footnote{The result is obtained calculating $j/R^3$ on the surface of the Earth at the equator.} From Eq. (\ref{eq:19}) we get
\begin{center}
$\delta\tau_{min}\sim 10^{-16}$\,s
\end{center}

The actual value of $\delta \tau_{min}$ depends of course on the adopted measurement strategy. If one thinks to directly measure times of flight using onboard atomic clocks the accuracy can hardly be in the order of $10^{-11}$ s. Optical interferometry, as the one applied in ring lasers \citep{ginger14}, would correspond to $\sim 10^{-17}$ s or better (one hundredth of a period or less). Finally, implementing the refined techniques in use for gravitational waves detection interferometers \citep{tae01,valli05} one could attain equivalent $\delta \tau_{min}$ values as low as $10^{-20}$ s. The latter approach leads to a value for the lowest detectable GM field
\begin{equation}
B_{MW}\geq\sim 10^{-27}\,\textrm{m}^{-1} \label{Bmin}
\end{equation}

Would this be enough? The effective angular momentum at the position of the Sun crucially depends not only on the mass of the Milky Way (be it baryonic or dark) but also on the way that mass is distributed and one has also to account for the fact that, unlike the gravito-electric interaction (ordinary gravitational attraction), also the matter distribution external to the orbit of the Sun plays a role. Anyway just to crudely have an idea of the orders of magnitude, if the Sun were at rest at the equator of a spinning homogeneous massive sphere and the peripheral velocity were Keplerian, the corresponding GM field would be
\begin{center}
$B_g=\frac{2}{5}\frac{m^{3/2}}{R_G^{5/2}}$
\end{center}
This, when using the value (\ref{Bmin}), points out a mass $m\sim 10^{16}$\,m. The visible mass of the Milky Way is $\sim 10^{14}$\,m and the estimated values for the dark mass are between $5$ and $10$ times bigger than the visible mass. All in all, and remembering that the model is unrealistic, we stay within one order of magnitude of the lowest attainable sensitivity with the best now available techniques.

\section{MODULATING THE SIGNAL}
\label{Section 6}

So far I have relied on the weakness of the effects, assuming that the contributions of different sources
could simply be summed. Once this has been assumed, the problem becomes: how can we discriminate the
various components? The galactic signal is indeed superposed to the ones due to the kinematical Sagnac
(dominant) and the terrestrial gravito-magnetism (but there are also the weaker terrestrial de Sitter term and
the solar gravito-magnetism). Lacking a well-defined idea of how big the galactic signal could be, we should
have to model, in the most accurate way, the other effects in order to subtract them from the raw data; the
residue would contain the information, but it would still be combined with different possible systematic errors.

To facilitate this difficult search, we may think to introduce a modulation of the interesting signal at a
predefined time rate. We could for instance build our Galilean triangle with two satellites on one orbital plane
and the third on a different orbital plane (the Galileo constellation is over three planes). If so, both the area of
the triangle and its inclination with respect to the axis of the Milky Way would periodically change in time; the
period would coincide with the orbital time of the constellation, i.e. $14$ h. The physical configuration is
sketched in Fig. 4.
\begin{figure}[h]
\label{figure4}
\begin{center}
\includegraphics*[width=16cm,angle=0]{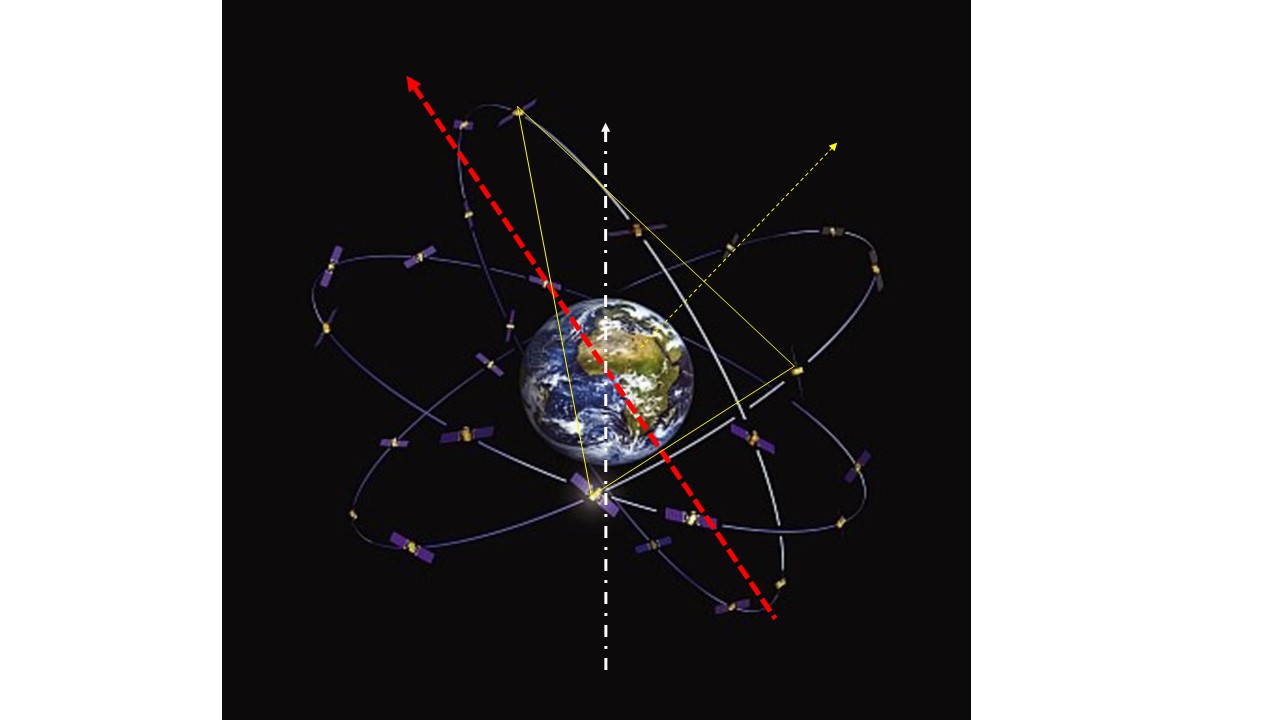}
\end{center}
\caption{Oscillating triangle based on Galileo
satellites located on two different orbital planes.}
\end{figure}
In the figure, the white arrow indicates the axis of the Earth; the red arrow is parallel to the axis of the Milky
Way; the yellow arrow is perpendicular to the plane of the triangle and oscillates in time. The relevant quantity is the product of the area of the triangle by the cosine of the inclination of its normal with respect to the axis of the galaxy. The combination of the change of the area and of the cosine, though being determined by plain Euclidean geometry, is not trivial. The result is not simply proportional to a sinusoid with a $14$ h period, but corresponds to a signal profile comprised of more than one Fourier component, which would be a peculiar signature for the sought for information. The "dance" of the satellites would become even more varied if one would choose to use three satellites each on a different orbital plane of the three of the constellation.

Adopting the above experimental configuration we should look, in the raw data, for periodical contributions at a base frequency corresponding to the $14$ h period but also having a non-trivial profile: they would most probably be originated by the Milky Way. Unfortunately there would however be a competitor left, which is the Sun: its GM field at the orbit of the Earth is also $\sim10^{-27}\,\textrm{m}^{-1}$ \citep{lag18,TA18} and the corresponding signal would be modulated in the same way as the one from the Milky Way. The possibility to discriminate the two contributions relies on the accuracy with which one knows the angular momentum of the Sun, which is not high.

\section{CONCLUSION}

As we have seen, asymmetric propagation of light offers an opportunity to measure gravito-magnetic fields, in
general, and to fix limits on the angular momentum of the Milky Way, including the contribution from its
rotating dark halo.
The Galileo constellation, activating the interlink communication among the satellites, offers suitable
configurations for such measurements; the more so if we think that involving communications among various sets of spacecraft would offer the opportunity to exploit a plurality of different oscillating triangles fit for measurements in the way presented in Sect. \ref{Section 6}. The required sensitivities are quite high but within the range of the best existing technologies. In order to say more, a specific analysis of the technological requirements and an in-depth discussion of the uncertainties is needed, both being beyond the scope of the present paper: suffices here to say that a reasonable suspect of feasibility in the near future is credible.

It is also interesting to remark that the measurement strategies could exploit synergies with terrestrial
experiments, based on ring lasers such as GINGER \citep{ginger17} (whose plane, moving around with the surface of the Earth,
oscillates daily with respect to the axis of the Milky Way) and missions in the inner solar system
(e.g. LISA and spacecraft located in the Lagrange points \citep{lag18}).

This search for weak effects of general relativity on a large scale looks interesting and promising: we should
take up the challenge.

\section{Acknowledgement}

This proposal has been inspired by the Galileo for Science collaboration and in particular by the group
below, which I wish to thank:

Lorenzo Casalino, Emiliano Fiorenza, Carlo Lefevre, David M. Lucchesi, Carmelo Magnafico, Roberto
Peron, Elisa Rosciano, Matteo Luca Ruggiero, Patrizia Sacco, Francesco Santoli, Francesco Vespe,
Massimo Visco.




\begin{thebibliography}{16}


\bibitem[Appourchaux et al.(2009)]{app09}
Appourchaux, T. et al. 2009, Astrodynamical Space Test of Relativity Using Optical
Devices I (ASTROD I)—A class-M fundamental physics mission proposal for Cosmic Vision 2015–2025, Exp. Astron. 23, 491–527.

\bibitem[Braxamaier et al.(2012)]{brax12}
Braxmaier, C., Dittus, H., Foulon, B. et al. 2012, Astrodynamical Space Test of Relativity using Optical Devices I (ASTROD I)— a class-M fundamental physics mission proposal for cosmic vision 2015–2025: 2010 Update, Exp. Astron. 34, 181–201.

\bibitem[Di Virgilio et al. (2014)]{ginger14}
Di Virgilio, A., Allegrini, M., Beghi, A. et al. 2014, A ring lasers array for fundamental physics, Comptes Rendus Physique, Elsevier, 15, 866-874.

\bibitem[Francis \& Anderson (2014)]{FA14}
Francis, C. \& Anderson, E. 2014, Two estimates of the distance to the Galactic Centre, Monthly
Notices of the Royal Astronomical Society {\bf 441} (2), 1105-1114.

\bibitem[Kafle, Sharma, Lewis \& Bland-Hawthorn (2014)]{KSLB14}
Kafle, Prajwal R., Sharma, Sanjib, Lewis, Geraint F. \& Bland-Hawthorn, Joss 2014, On the
Shoulders of Giants: Properties of the Stellar Halo and the Milky Way Mass Distribution. The
Astrophysical Journal {\bf 794} (1), 17.

\bibitem[Ruggiero \& Tartaglia (2002)]{RT2}
Ruggiero, M.L. \& Tartaglia, A. 2002, Gravitomagnetic effects, Nuovo Cimento B 117, 743-767.

\bibitem[Ruggiero \& Tartaglia (2019)]{RT19}
Ruggiero, M.L. \& Tartaglia, A. 2019, Test of gravitomagnetism with satellites around the Earth,
Eur. Phys. J. Plus 134, 205.

\bibitem[Tartaglia (2010)]{TA10}
Tartaglia, A. 2010, Emission coordinates for the navigation in space, Acta Astronautica 67, 539-
545.

\bibitem[Tartaglia(2018)]{TA18}
Tartaglia, A. 2018, Dark angular momentum of the galaxy, Int. J. Mod. Phys. D, {\bf 27}, 14, 1847012-1.

\bibitem[Tartaglia, Di Virgilio, Belfi, Beverini, \& Ruggiero (2017)]{ginger17}
Tartaglia, A., Di Virgilio, A., Belfi, J., Beverini, N.
\& Ruggiero, M.L. 2017, Testing general relativity by means of ring lasers, Eur. Phys. J.
Plus 132, 73.

\bibitem[Tartaglia et al. (2017)]{TA17}
Tartaglia, A, Ruggiero, M.L., Delle Monache, G. et al. 2017, Inverse Relativistic
Positioning for G4S, in Proc. 6th "International
Colloquium on Scientific and Fundamental Aspects of GNSS/Galileo", 25-27 October 2017, Valencia.

\bibitem[Tartaglia et al.(2018)]{lag18}
Tartaglia, A., Lorenzini, E.C., Lucchesi, D., Pucacco, G., Ruggiero, M.L. \& Valko,P. 2018,
How to use the Sun–Earth Lagrange points for fundamental physics and navigation, Gen. Relativ.
Gravit. 50, 9.

\bibitem[Tartaglia, Ruggiero \& Capolongo (2011)]{TRC11}
Tartaglia, A., Ruggiero, M.L. \& Capolongo, E. (2011). A null frame for spacetime positioning by
means of pulsating sources, Advances in Space Research 47, 645-653.

\bibitem[Tinto, Armstrong \& Estabrook (2001)]{tae01}
Tinto, M., Armstrong, J.W. \& Estabrook, F.B. 2001, Discriminating a gravitational-wave background from
instrumental noise using time-delay interferometry, Class. Quantum Grav. 18, 4081-4086.

\bibitem[Vallisneri (2005)]{valli05}
Vallisneri, M. 2005, Geometric time delay interferometry, Phys. Rev. D, 72, 042003.

\bibitem[Wechsler \& Tinker (2018)]{WT18}
Wechsler, R.H. \& Tinker, J.L. 2018, The Connection between Galaxies and their Dark
Matter Halos, Annual Review of Astronomy and Astrophysics 56, 435-487.

\end{thebibliography}
\end{document}